\documentclass[longbibliography,prb]{revtex4-2}

\usepackage{amsmath}
\usepackage{hyperref}
\usepackage{graphicx}

\usepackage{xcolor}

\newcommand{\VEC}[1]{{\boldsymbol{ #1}}}

\newcommand{\ie}{{\it i.e.}}

\newcommand{\half}{\frac{1}{2}}
\newcommand{\Fig}{{Fig.}}
\newcommand{\Eq}{{Eq.}}

\newcommand{\be}{\begin{equation}} \newcommand{\ee}{\end{equation}}
\newcommand{\bea}{\begin{eqnarray}} \newcommand{\eea}{\end{eqnarray}}

\newcommand{\FP}{Fokker-Planck}
\newcommand{\LLG}{Landau-Lifshitz-Gilbert}
\newcommand{\origHz}{H_{\rm k}^{\rm{eff}}}
\newcommand{\origHy}{H_{\rm{k,in}}}
\newcommand{\origHx}{H_{\rm{in}}}

\newcommand{\gammazero}{\gamma}

\begin{document}
\title{
In-plane dominant anisotropy stochastic magnetic tunnel junction for probabilistic computing: A Fokker-Planck study
}

\author{Chee Kwan Gan}
\email{ganck@ihpc.a-star.edu.sg}
\affiliation{Institute of High Performance Computing (IHPC), Agency for Science, Technology and Research (A*STAR), 1 Fusionopolis Way, \#16-16 Connexis, Singapore 138632, Republic of Singapore}
\author{Bingjin Chen}
\email{chen\_bingjin@ihpc.a-star.edu.sg}
\affiliation{Institute of High Performance Computing (IHPC), Agency for Science, Technology and Research (A*STAR), 1 Fusionopolis Way, \#16-16 Connexis, Singapore 138632, Republic of Singapore}
\author{Minggang Zeng}
\email{zeng\_minggang@i2r.a-star.edu.sg}
\affiliation{Institute of Infocomm Research (I2R), Agency for Science, Technology and Research (A*STAR), 1 Fusionopolis Way, \#21-01 Connexis, Singapore 138632, Republic of Singapore}

\date{Aug 23, 2022}

\begin{abstract}
Recently there is considerable interest to realize efficient and low-cost 
true random number generators (RNGs) for practical applications.
One important way is through the use of bistable magnetic tunnel junctions (MTJs).
Here we study the magnetization dynamics 
of an MTJ, with a focus to realize efficient random bit generation under the assumption that
the orientation dependence of the 
energy of the nanomagnet is described by two perpendicular in-plane anisotropies. 
We find that a high rate of random bit generation is achievable
away from the pure easy-axis situation by tuning a single parameter $H_z$ so that
it is either (a) toward a barrierless-like single easy plane 
situation when $H_z$ reduces to zero, or (b) toward a stronger easy plane situation
when $H_z$ becomes increasingly negative where transitions between low energy states 
are confined in the stronger easy plane that contains the saddle points.
We find that the MTJs maintain their fast magnetization dynamical characteristics even in the presence of a magnetic field.
Our findings provide a valuable guide to achieving
efficient generation of probabilistic bits for applications in probabilistic computing.
\end{abstract}


\maketitle
\section{Introduction}

The next-generation computing paradigms such as 
probabilistic computing \cite{McGoldrick22v13,Kaiser21v119,Finocchio21v521,Talatchian21v104,Liao21v27,Wu22v18,Aadit22v5,Zink22v8,Camsari21v15,Debashis20v10,Debashis20v101,Hassan19v10,Pyle18v12,Zink18v124,Roy18v123,Suh15v117,Bapna17v111,Cai23v129,Chowdhury23v9,Zink23v9,Cai23v66}
and quantum computing\cite{Feynman82v21,Nelsen2016-book,Albash18v90}
hold great promise for
solving problems that are very difficult or impossible with
the traditional von Neumann computing technologies.
In probabilistic computing, a problem may be solved by encoding the solution into the 
physics of the model,\cite{Bian10-Dwave,Biamonte08v77,Whitfield12v99}
where the phase space is to be sampled efficiently using a physical device 
to accelerate computing. 
These include the integer factorization problem\cite{Borders19v573}, the traveling 
sale person problems\cite{Sutton17v7}, 
the invertible Boolean logic problem\cite{Camsari17v7},
the maximum satisfiability problem\cite{Grimaldi22v17}, and the max-cut problem.\cite{Sutton20v8}
The probabilistic bits (p-bits) could be used to perform 
basic arithmetic functions such as 32-bit adder or subtractor\cite{Faria17v8}.
These p-bits could also be interconnected to build correlated p-circuits to implement
useful invertible AND gates.\cite{Camsari17v38}
The parallel versus antiparallel resistance ratio of the stochastic 
MTJs could be harnessed by inverters and amplifiers.\cite{Camsari21v15,Camsari17v38}
Superparamagnetic tunnel junctions have been demonstrated to achieve low-energy and
high-quality random bit generation.\cite{Camsari17v38,Vodenicarevic17v8,Parks18v8}
The MTJs could also be used in neural computing\cite{Deng20v41,Misra22vX,Liu22vX} 
to simulate stochastic neurons for a hardware implementation of a restricted Boltzmann machine.

Most of the applications mentioned above rely on the availability of low-cost and effective RNGs.\cite{Chen22v18,Rehm23v19,Liu22v8,Debashis20v101,Debashis22v13,Jenkins19v9}
High quality random numbers are also needed in many other applications such as the cryptographic systems and Monte Carlo simulations.
Recently, the telegraphic switching characteristics of MTJs
have been experimentally demonstrated.\cite{Hayakawa21v126,Safranski21v21,Sun23v107,Kim22v12}
In this work, we study the magnetization dynamics of a nanomagnet where the energy expression\cite{Hayakawa21v126} 
contains two perpendicular easy planes with the anisotropy parameters $H_z$ and $H_y$. 
We use the \FP{} equation 
approach\cite{Taniguchi13v88,Taniguchi12v85,Brown63v130,Li04v69,Cheng06v96,Apalkov05v72,Butler12v48,Holubec19v99,Liu21v11,Das21v68,Kanai21v103} to calculate the relaxation time that is a good measure of the random switching rate deduced 
from the \LLG{} approach.\cite{Kanai21v103} 
We find that it is possible to achieve an enhanced telegraphic switching rate
as the parameter $H_z$ is reduced to zero (so that the transition becomes barrierless in the limit).
This provides an alternative approach to increase the switching rate since it has been 
shown that it is also possible to increase the switching rate
by increasing the magnitude of $H_z$.\cite{Hayakawa21v126,Kanai21v103}
The application of a magnetic field gives rise to a sigmoid function behavior,\cite{Kim22v12,Kobayashi21v119}
where the fast switching rate observed in the absence of magnetic field is largely maintained. 
Good tunability of the stochasticity of MTJs may be achieved by other means such the spin-transfer torque or the spin-orbit torque.\cite{Lu22v8,Shao21v57,Ostwal19v10,Xie17v64,Liu12v109,Koch04v92}
This paper is organized as follows. Section~\ref{sec:method} introduces 
the basic equations governing the magnetization dynamics and outlines 
the methodology used to solve the \FP{} equation.
Section~\ref{sec:results} contains 
the analysis of the results of the simulations. 
The summary and conclusions are found in Section~\ref{sec:summary}.

\section{Methodology}
\label{sec:method}

We study the magnetization dynamics of an MTJ 
using the stochastic \LLG{} (LLG) equation\cite{Brown63v130} with the Langevin (fluctuation field) term as given by
\be
\frac{d\VEC{M}}{dt} = \gammazero \VEC{M} \times \left[  - \frac{\partial U }{\partial \VEC{M}} - \eta \frac{d \VEC{M}}{d t}  + \VEC{h}(t)\right]
\label{eq:llg}
\ee
where $\VEC{M}$, $t$, $\gammazero$, $U$, $\eta > 0$ are the magnetization, time, gyromagnetic ratio, energy density, and dissipation constant, respectively.
The Gilbert damping constant is given by $\alpha= \eta |\gammazero| M_s$, where $M_s$ is the saturation magnetization.
The components of the random field $\VEC{h}(t)$, \ie, $h_i(t)$, $i=1,2,3$, satisfy the conditions $\langle h_i(t)\rangle = 0$, 
and $\langle h_i(t) h_j(t+s) \rangle = \mu \delta_{ij} \delta(s)$. Here $\langle x \rangle$ means the statistical 
average of $x$, $\mu = 2k_B T\eta/V$, where $T$ is the temperature and $V$ the volume of the magnet.
It is common to describe 
$\VEC{M}$ by a unit vector $ \VEC{m} = \VEC{M}/M_s = (m_x, m_y, m_z) = (\sin\theta \cos\varphi, \sin\theta \sin\varphi, \cos\theta)$.

The energy density $U$ for the nanomagnet\cite{Hayakawa21v126} is given by
\bea
U = \frac{E}{V} &=& \frac{1}{2} \mu_0 M_s \left( \origHz{} + \origHy{} \sin^2 \varphi   \right) \sin^2\theta   \nonumber
\\ && -\mu_0 M_s \origHx{} \sin\theta \cos\varphi
\label{eq:U}
\eea
where $E$ is the energy of the magnet.
Notice that $\origHz{}<0 $ describes an in-plane anisotropy
field. However, $\origHy{} > 0  $ also describes an in-plane anisotropy field.
We assume that an applied magnetic field of magnitude $\origHx{}$ is pointing in the positive $x$ direction.

In terms of the components of $\VEC{m}$, \Eq~\ref{eq:U} can be rewritten as
\bea
E &=& \frac{1}{2} \mu_0 M_s V \left[ \origHz{} (1-m_z^2) 
- \origHy{} (1-m_y^2) \right]     \nonumber
\\&& + \frac{1}{2} \mu_0 M_s V \origHy{}  -\mu_0 M_s V \origHx{} m_x
\label{eq:energy}
\eea
If we introduce $H_z = \origHz{}$,  $H_y = -\origHy{}$, and $H_x = \origHx{}$, we obtain a more symmetric form for \Eq~\ref{eq:energy} where
\bea
E &=& \frac{1}{2} \mu_0 M_s V \left[ H_z (1-m_z^2) + H_y (1-m_y^2) \right]  \nonumber
\\&& - \frac{1}{2} \mu_0 M_s V H_y -\mu_0 M_s V H_x m_x
\label{eq:symE}
\eea
The advantage of \Eq~\ref{eq:symE} is that the signs of $H_z$ and $H_y$  have the  same
physical meaning, \ie, $H_z < 0$ is the in-plane anisotropy parameter in the $xy$ plane while
$H_y < 0$ is the in-plane anisotropy parameter in the $xz$ plane.
We note that if both the in-plane anisotropy parameters are the same, \ie, $H_z = H_y = H < 0$, then
\Eq~\ref{eq:symE} reduces to
\bea
E &=& -\frac{1}{2} \mu_0 M_s V H (1-m_x^2) + \frac{1}{2} \mu_0 M_s V H   \nonumber
\\&& -\mu_0 M_s V H_x m_x
\label{eq:E-of-easy-axis}
\eea
which corresponds to an easy $x$ axis anisotropy situation, since 
the intersection of two equivalent perpendicular easy planes gives rise to
an easy axis.

One way to study the dynamics of the magnetic system is to directly
integrate the equation of motion\cite{Donahue16,GarciaPalacios98v58,Tiwari14v104}  as described by
\Eq~\ref{eq:llg}. Another way is to 
study its associated 
\FP{} equation\cite{Brown63v130}
\bea
\frac{\partial W}{\partial t} 
&=& \frac{1}{\sin\theta} \frac{\partial}{\partial \theta} 
\left\{   \left( h'  \sin\theta \frac{\partial U}{\partial \theta} - g'\frac{\partial U}{\partial \varphi}\right) W  + k' \sin\theta \frac{\partial W}{\partial \theta}  \right\}  \nonumber
\\ &+& \frac{1}{\sin\theta} \frac{\partial}{\partial \varphi} \left\{  \left(  g'\frac{ \partial U }{ \partial \theta }  + \frac{h'}{\sin\theta} \frac{ \partial U  }{ \partial \varphi  } \right) W + \frac{k'}{\sin\theta} \frac{\partial W}{\partial \varphi}   \right\}
\label{eq:FPE-W}
\eea
where $h' = \frac{\eta \gammazero^2}{1 + \alpha^2} $, $g' = \frac{\gammazero}{M_s(1+ \alpha^2)}$,
and $k' = \frac{ \mu \gamma^2  }{  2(1 + \alpha^2)  }$.
Here $ W(\theta,\varphi,t) \sin\theta d\theta d\varphi$ is the probability of finding $\VEC{M}$, 
at time $t$, in the interval
defined by  $(\theta,\theta+d\theta) $ and $(\varphi,\varphi+d\varphi)$.

To incorporate the boundary conditions involving $\theta$ more naturally, 
we introduce a variable $ u(\theta,\varphi,t)  = W(\theta,\varphi,t) \sin \theta$ that guarantees 
that $u  = 0$ when $\theta = 0$ or $\theta = \pi$ (which corresponds to a
hard boundary condition\cite{Holubec19v99}).
This results in an alternative form of the \FP{} equation as given by
\bea
\frac{\partial u }{\partial t}
&=&
\frac{\partial}{\partial \theta} \left[ h' \frac{\partial U}{\partial \theta} u  - \frac{g'}{\sin\theta} \frac{\partial U}{\partial \varphi} u   - k' \frac{\cos\theta}{\sin \theta}  u   \right] \nonumber
\\&& 
+ \frac{\partial}{\partial \varphi} \left[  \frac{g'}{\sin\theta} \frac{ \partial U}{\partial \theta} u  + \frac{h'}{\sin^2\theta} \frac{ \partial U  }{  \partial \varphi }  u  \right]   \nonumber
\\ && + k' \frac{ \partial^2  }{ \partial \theta^2  } u 
+ \frac{k'}{\sin^2\theta} \frac{ \partial^2    }{ \partial \varphi^2  } u 
\label{eq:FPE-P}
\eea
For the initial condition,
we assume that $\VEC{M}$ points preferentially in the $-x$ direction and is
approximated by a 2D Gaussian function
\bea
u (\theta,\varphi,t=0) &=& \frac{1}{2\pi \sigma_\theta \sigma_\varphi} 
\exp\left[ - \frac{ (\theta- {\overline \theta})^2  }{2\sigma_\theta^2}\right] \nonumber
\\&&
 \exp\left[ - \frac{ (\varphi- {\overline \varphi})^2  }{2\sigma_\varphi^2}\right]
\label{eq:Pinit}
\eea
where ${\overline \theta} = \pi/2$ and ${\overline \varphi} = \pi$. The variances
$\sigma_{\theta}^2$ and $\sigma_{\varphi}^2$ should be chosen in such a way that 
they are reasonably 
small to describe a localized distribution but large enough
to avoid numerical instabilities.
The initial condition of \Eq~\ref{eq:Pinit} preserves the symmetric roles of $H_z $ and $H_y$ in \Eq~\ref{eq:symE}, therefore
we should obtain essentially the same dynamics if we swap the 
values of $H_z $ and $H_y$ in the simulations.

The split operator technique\cite{Altaai20v1591,Butt21v180,Press2002-book}
is used to evolve the variable $u $ in \Eq~\ref{eq:FPE-P}.
In this work, we find that 
the simple Forward Time Centered Space (FTCS) scheme (Appendix \ref{sec:Diffusion}) is 
accurate enough to handle the diffusion terms, \ie,
the third and fourth terms on the right hand side of \Eq~\ref{eq:FPE-P} involving 
the second derivatives of $u $ with respect to $\theta$ and $\varphi$, respectively. 
The results of our 
FTCS scheme are compared to those that are based on the robust Crank-Nicholson scheme 
but no significant deviation is obtained.
For the drift terms (\ie, the first and second terms on the right hand side
of \Eq~\ref{eq:FPE-P}), the Lax-Friedrichs (Appendix \ref{sec:Drift})  scheme is inappropriate since 
the dynamics is too dissipative. Fortunately, 
both the Lax-Wendroff or the MacCormack schemes (Appendix \ref{sec:Drift}) 
do not have the serious dissipative issue as compared to the Lax-Friedrichs scheme
and they both give essentially identical dynamics. 

\section{Results and discussions}
\label{sec:results}

\begin{figure}
\centering\includegraphics[width=9.2cm,clip]{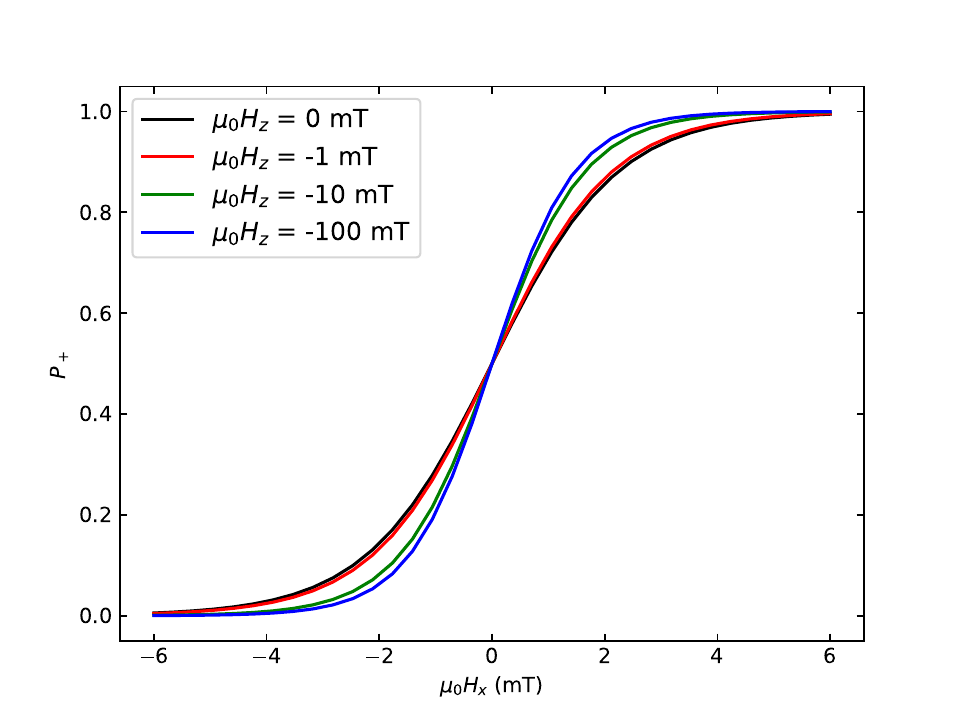}
\caption{$P_+$ as a function of applied magnetic field
$\mu_0 H_x$. $P_+$ displays increasingly localized features as 
we increase $\mu_0 H_z$ in the sequence of $0, -1, -10, -100$~mT.  We use $\mu_0 H_y = -10$~mT in all calculations. The subroutine 
{\tt dblquad} from the python package scipy is used for numerical integrations.
}
\label{fig:2023-04-18-Pplus-vary-Hx-for-diff-Hz.pdf}
\end{figure}

First we study the magnetization dynamics of the nanomagnet
in the absence of an applied magnetic field, \ie,  $\mu_0 H_x = 0$~mT. 
In our simulations, we use the parameters adopted 
in Ref.[\onlinecite{Kanai21v103}] where
$\gammazero =  -1.7588 \times 10^{11}~\rm{T}^{-1}\cdot\rm{s}^{-1}$, 
$M_s = 1.114 \times 10^6~\rm{A}\cdot\rm{m}^{-1}$, and $\alpha = 0.02$. 
The nanomagnet has a diameter of $60$~nm and a thickness of $1$~nm.
The temperature $T=300$~K is used throughout. 
A hard boundary condition is assumed for $u$ in $\theta$, where $0  \le \theta \le \pi$,
but a periodic boundary condition is assumed for $u$ in $\varphi$, where $-\pi/2 \le \varphi < 3\pi/2$.
The number of divisions in the $\theta$ range, $N_\theta = 100$, 
while the number of divisions in the $\varphi$ range, $N_\varphi=200$. These are found to be sufficient to
ensure convergence of the results.
The values of $\sigma_{\theta} = \sigma_{\varphi} = \pi/18$ are used.
At time $t$, the probability to find the magnetization $\VEC{M}$ pointing in the $+x$ hemisphere is given by
\be
P_s(t) = \int\int_{m_x > 0}  u (\theta,\varphi,t) d\theta d\varphi
\ee

At large enough $t$, $P_s(t)$ attains an equilibrium value as given by
\be
P_+  = \frac{1}{Z} \int_{-\pi/2}^{\pi/2} d\varphi \int_0^{\pi}d\theta \  e^{-E/k_BT} \sin\theta 
\ee
where $Z$ is the partition function as given by
\be
Z  = \int_{-\pi/2}^{3\pi/2} d\varphi \int_0^{\pi}d\theta \  e^{-E/k_BT} \sin\theta 
\ee

\begin{figure}
\centering\includegraphics[width=9.2cm,clip]{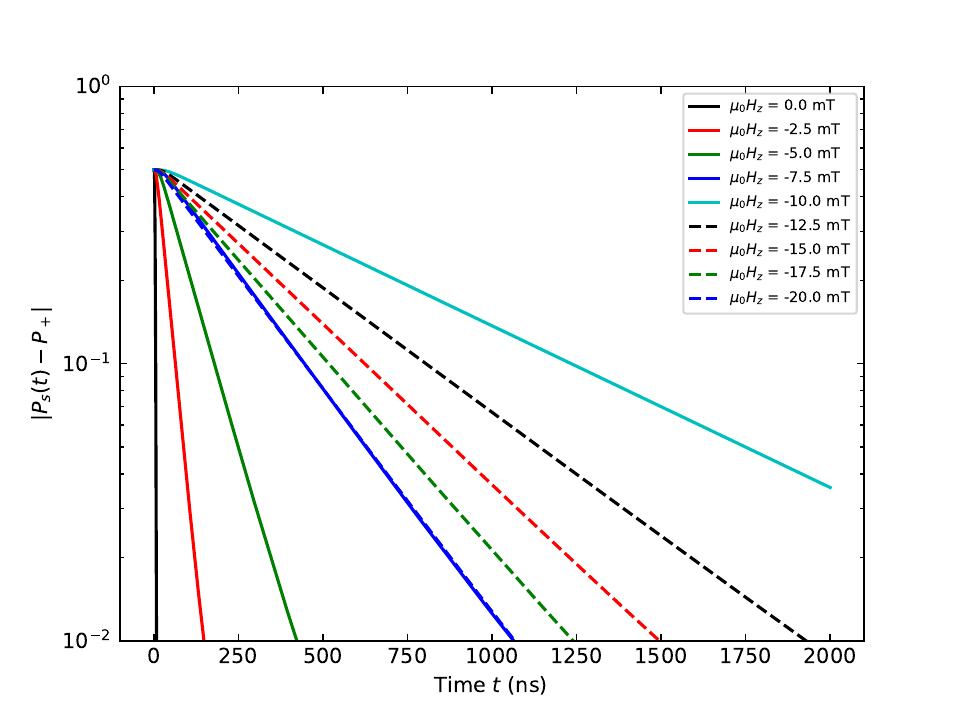}
\caption{The approach of the probability $P_s(t)$ toward $P_+ = \frac{1}{2} $ as a function of time $t$ for
$\mu_0 H_z = 0.0$, $-2.5$, $-5.0$, $-7.5$, $-10.0$, $-12.5$, $-15.0$, $-17.5$, $-20.0$~mT
with $\mu_0 H_y = -10$~mT. No magnetic field is present (\ie, $\mu_0 H_x = 0$~mT). 
}
\label{fig:truncated-2023-01-26-Ps-vary-Hz-mu0Hy-minus10-mu0Hx-0.pdf}
\end{figure}

In the absence of an applied magnetic field, \ie, $\mu_0 H_x=0$~mT, there is an equal probability to find $\VEC{M}$ in the $+x$ or $-x$ hemispheres
and therefore
$P_+ = \frac{1}{2}$. However, when $H_x > 0$, $\VEC{M}$ will preferably reside  in the $+x$ hemisphere
due to a lower energy and we expect $P_+ > \frac{1}{2}$, which is confirmed by the results
shown in \Fig~\ref{fig:2023-04-18-Pplus-vary-Hx-for-diff-Hz.pdf}.
The $P_+$ versus $\mu_0 H_x$ curves for several chosen $\mu_0 H_z$ values are seen to be sigmoid functions.

\begin{figure}
\centering\includegraphics[width=14.2cm,clip]{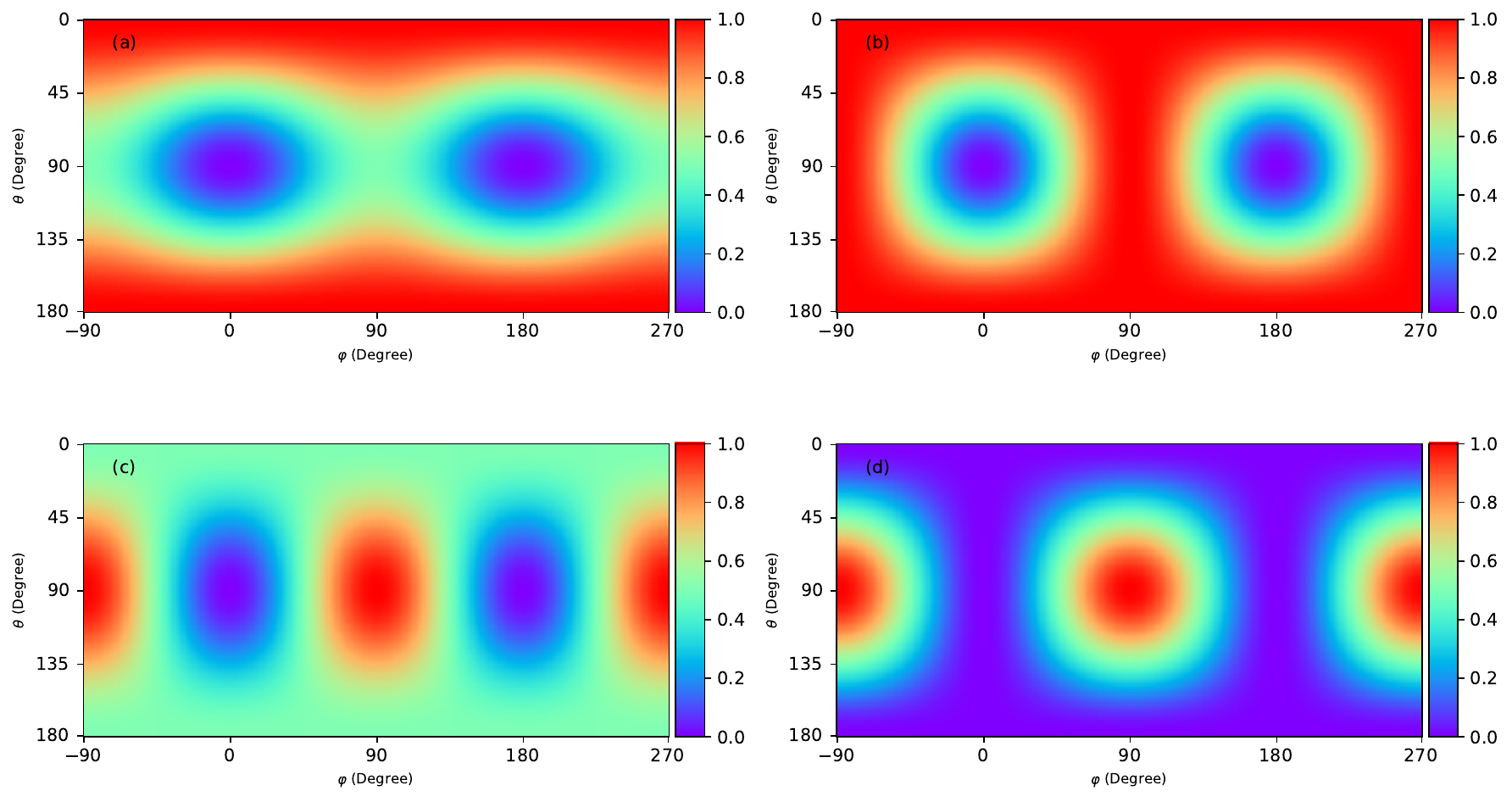}
\caption{
The energy profile when 
(a) $\mu_0 H_z =-20.0$~mT, 
(b) $\mu_0 H_z =-10.0$~mT, 
(c) $\mu_0 H_z =-5.0$~mT, 
and
(d) $\mu_0 H_z =0.0$~mT. In all cases $\mu_0 H_y = -10$~mT and $\mu_0 H_x = 0$~mT.
The energies have been scaled so that they range between $0$ and $1$.
}
\label{fig:2023-05-05-Energy-contours.pdf}
\end{figure}

\Fig~\ref{fig:truncated-2023-01-26-Ps-vary-Hz-mu0Hy-minus10-mu0Hx-0.pdf} shows the exponential approach of $P_s(t)$ 
toward the limiting value of $P_+ = \frac{1}{2}$ for several values of $\mu_0 H_z$. 
This allows us to determine the relaxation time $\tau $ from a linear fit to $ |P_s(t) - P_+| = A e^{-t/\tau}$ for some  
constant $A$. 
It should be pointed out that $\tau$ determined from the \FP{} approach can be also be
estimated from the magnetization trajectories evolved according to
the \LLG{} equation, as demonstrated in Ref.[\onlinecite{Kanai21v103}].
We obtain $\tau = 742.2$, $483.0$, $373.7$, $313.5$, and $268.3$~ns for $\mu_0 H_z = -10.0$, $ -12.5$, $-15.0$, $-17.5$, and $-20.0$~mT, respectively.
The value of $\tau =742.2$~ns when $\mu_0 H_z = \mu_0 H_y = -10$~mT agrees very well with that of 
 $\tau \sim 770$~ns using a rather similar method\cite{Kanai21v103}.
Since $H_z = H_y$, it describes the easy $x$ axis situation
and $P_s(t)$ approaches $P_+$ rather slowly compared to all other values of $\mu_0 H_z$. 
In the easy axis situation, there is
an energy barrier on the $yz$ plane (\ie, $\varphi = \pm \pi/2$) separating the two lowest energy states as shown in \Fig~\ref{fig:2023-05-05-Energy-contours.pdf}(b).
The shorter relaxation time obtained with very large negative $\mu_0 H_z $ values is due to concentrated transitions across
two low-energy saddle points $(\theta,\varphi) = (\pi/2, \pm \pi/2)$ [see \Fig~\ref{fig:2023-05-05-Energy-contours.pdf}(a)]
which lie in the $xy$ easy plane to minimize
the energy penalty while incurring a fixed penalty due to a deviation from the $xz$ easy plane\cite{Kanai21v103}.


\begin{figure}
\centering\includegraphics[width=9.2cm,clip]{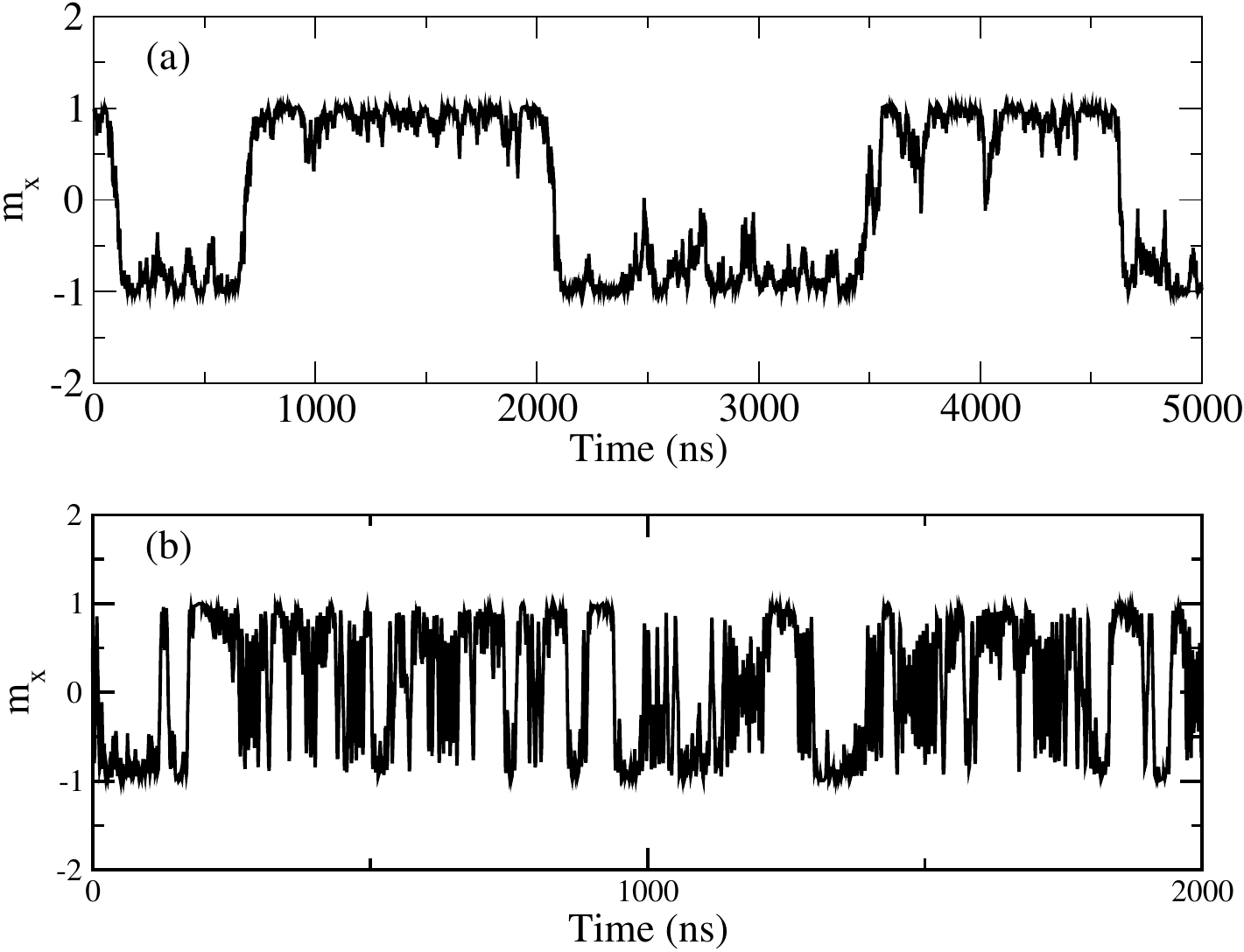}
\caption{The $x$ component of the magnetization 
of the nanomagnet
simulated by the \LLG{} equation 
with the Langevin term at $300$~K 
for (a) $\mu_0 H_z = -10$~mT and (b) $\mu_0 H_z = -2.5$~mT. 
We have used 
$\mu_0 H_y = -10$~mT.  There is no applied magnetic field.
}
\label{fig:Telegraphic-Hz-minus10-vs-Hz-minus2.5.pdf}
\end{figure}

When $\mu_0 H_z $ is decreased from $-10.0$~mT toward $0.0$~mT, we find that the 
relaxation time is also reduced compared to that of the easy-axis situation.
We find $\tau = 742.2$, $ 262.0$, $102.2$, $35.6$, and $0.91$~ns for $\mu_0 H_z = -10.0$, $ -7.5$, $-5.0$, $-2.5$, and $0.0$~mT, respectively.
In the limit when $\mu_0 H_z = 0$~mT, $\VEC{M}$ could traverse within 
the $xz$ easy plane without any energy penalty and there is no barrier to move between
the positive $x$ and negative $x$ 
hemispheres as long as $\varphi=0$ or $\pi$ [see \Fig~\ref{fig:2023-05-05-Energy-contours.pdf}(d) for the energy profile].
In \Fig~\ref{fig:Telegraphic-Hz-minus10-vs-Hz-minus2.5.pdf} we show a representative case where $\mu_0 H_z = -2.5$~mT
deliver a higher random switching rate than that for the case where $\mu_0 H_z = -10$~mT. 

\begin{figure}
\centering\includegraphics[width=9.2cm,clip]{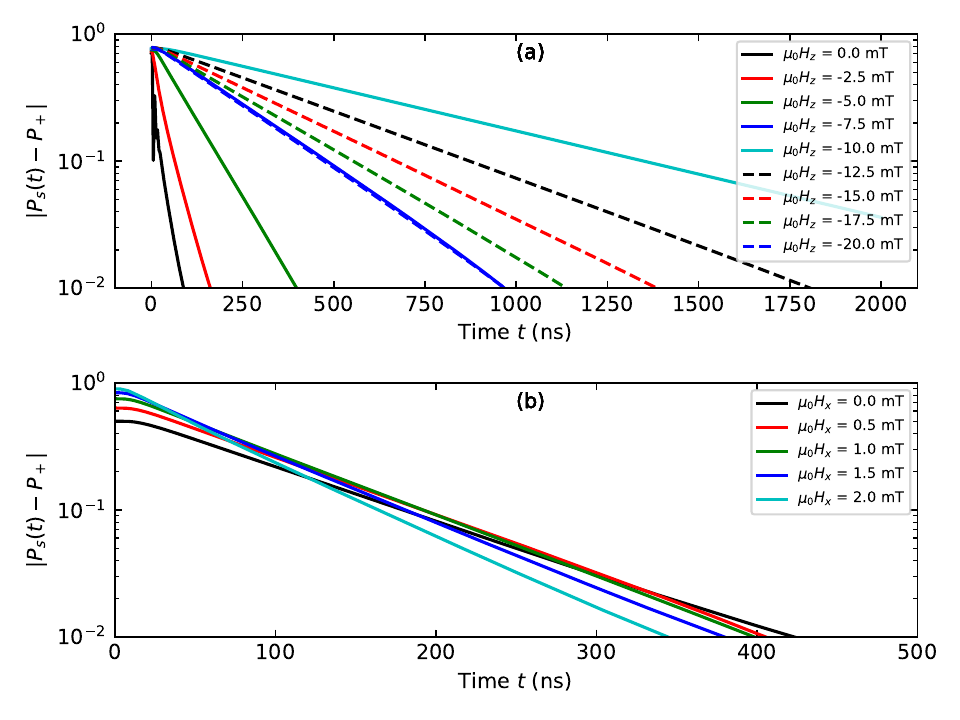}
\caption{(a) The approach of the probability $P_s(t)$ toward its respective $P_+$ as a function of time $t$ for
$\mu_0 H_z = 0.0$, $-2.5$, $-5.0$, $-7.5$, $-10.0$, $-12.5$, $-15.0$, $-17.5$, and $-20.0$~mT
with $\mu_0 H_y = -10$~mT. The applied magnetic field $\mu_0 H_x = 1$~mT. 
(b) 
The approach of the probability $P_s(t)$ toward its respective $P_+$
as a function of time $t$ for applied magnetic fields of
$\mu_0 H_x = 0.0$, $0.5$, $1.0$, $1.5$, and $2.0$~mT.  $\mu_0 H_z = -5$~mT and $\mu_0 H_y = -10$~mT are chosen.
}
\label{fig:combined-truncated-2023-04-17-Ps-vary-Hz-mu0Hy-minus10-mu0Hx-1.pdf}
\end{figure}

Next we discuss the magnetization dynamics in the presence of an applied magnetic field.
When the magnetic field of $\mu_0 H_x = 1$~mT is present,
\Fig~\ref{fig:combined-truncated-2023-04-17-Ps-vary-Hz-mu0Hy-minus10-mu0Hx-1.pdf}(a)
shows the magnetization dynamics is rather similar to that shown in \Fig~\ref{fig:truncated-2023-01-26-Ps-vary-Hz-mu0Hy-minus10-mu0Hx-0.pdf}
where the magnetic field is absent. 
The relaxation time is seen to decrease when $\mu_0 H_z$ changes from $-10.0$~mT toward $-20.0$~mT, where
we have $\tau$ $= 640.7$, $411.9$, $315.8$, $259.6$, and $221.1$~ns for $\mu_0 H_z = -10.0$, $ -12.5$, $-15.0$, $-17.5$, and $-20.0 $~mT, respectively. 
The relaxation time also decreases when $\mu H_z$ changes from $-10.0$~mT toward $0.0$~mT, 
where $\tau =  640.7$, $223.0$, $90.3$, $38.6$, and $24.2$~ns for $\mu_0 H_z = -10.0$, $ -7.5$, $-5.0$, $-2.5$, and
$0.0$~mT, respectively.
\Fig~\ref{fig:combined-truncated-2023-04-17-Ps-vary-Hz-mu0Hy-minus10-mu0Hx-1.pdf}(b) 
shows that $P_s(t)$ approaches to its respective limiting value of $P_+$ in a similar manner for various values of
$\mu_0 H_x$ for a typical setting of $\mu_0 H_z = -5$~mT and $\mu_0 H_y = -10$~mT. 
This shows that the same switching characteristics could be preserved in the presence of a magnetic field.

\section{Summary and conclusions}
\label{sec:summary}

We have studied the magnetization dynamics of a magnetic tunnel junction using
the efficient \FP{} approach. A time-splitting operator method 
was used to integrate the four different terms in the \FP{} partial differential equation (in
three independent variables, \ie, the polar angle $\theta$, the azimuthal angle $\varphi$, and the time $t$)
as derived from the stochastic \LLG{} equation. 
We identified the energy expression as a two-perpendicular easy planes situation as characterized by the anisotropy parameters $H_z$ and $H_y$.
Without loss of generality, by fixing the value of $H_y$, we studied the effect of $ H_z$ on the magnetization dynamics
and deduced the relaxation times that determine the random switching rates.
We found that when $H_z = H_y$ the dynamics is the slowest since it corresponds to an easy axis situation. 
Decreasing $H_z$ (\ie, increasing the strength of the $xy$ easy plane) results in an enhanced
random switching rate due to a strong constraining effect in the $xy$ plane. 
On the other hand, decreasing the magnitude of $\mu_0 H_z$ toward  $0.0$~mT
also results in a fast random switching dynamics where a barrierless-like situation is approached.
The application of the magnetic field results in a sigmoid function in the $P_+$ versus $\mu_0 H_x$
curve (where $P_+$ is the equilibrium probability of finding the magnetization in the $+x$ hemisphere)
and it does not affect random switching characteristics when there is no magnetic field. 
Our findings provide a valuable guide 
to achieving fast random switching in a true random number generator
for applications in probabilistic computing.

\section{Acknowledgment}
This work is supported by Agency for Science, Technology and Research (A*STAR) under
Career Development Fund (Project No. C210812054).

\bibliography{master-references}

\appendix

\section{Solving the diffusion equation}
\label{sec:Diffusion}

We consider the diffusion equation
\be
\frac{\partial u}{\partial t} = A(x) \frac{\partial }{\partial x} \left[
 D(x) \frac{\partial u}{\partial x}\right]
\ee
The explicit Forward Time Centered Space (FTCS) scheme is given by
\bea
\frac{ u_i^{n+1} - u_i^n}{\Delta t}  &=&
\frac{A_i}{(\Delta x)^2} \left[  D_{i+\half} (u_{i+1}^n - u_i^n) 
\right.
\\ & & -
\left.
D_{i-\half} (u_i^n - u_{i-1}^n)   \right]
\eea
where $A_i = A(x_i) $, $D_{i+\half} = D(x_{i+\half})$, etc.

\section{Solving the drift equation}
\label{sec:Drift}

We consider the drift equation
\be
\frac{\partial u}{\partial t} = - \frac{\partial F}{\partial x}
\ee

The Lax-Friedrichs scheme is given by
\be
u_{j}^{n+1} = \frac{1}{2} (u_{j+1}^n + u_{j-1}^n) - \frac{\Delta t}{2 \Delta x}
( F^n_{j+1} - F^n_{j-1})
\ee

The Lax-Wendroff scheme is given by
\bea
u_{i+\half}^{n+\half} &=& \half( u_{i+1}^n + u_i^n) 
\\ && - \frac{\Delta t}{2 \Delta x} ( F(u_{i+1}^n) - F(u_i^n))
\\ u_{i-\half}^{n+\half} &=& \half( u_{i}^n + u_{i-1}^n) 
\\ && - \frac{\Delta t}{2 \Delta x} ( F(u_{i}^n) - F(u_{i-1}^n))
\eea
and then
\be
u_i^{n+1} = u_i^n - \frac{\Delta t}{\Delta x} \left[ F(u_{i+\half}^{n+\half})
- F( u_{i-\half}^{n+\half})  \right]
\ee

Finally the MacCormack scheme is 
given by

\be
u_j^p = u_j^n - \frac{\Delta t}{\Delta x} \left(  F_{j+1}^n - F_j^n \right)
\ee
and
\be
u_j^{n+1} = \half( u_j^n + u_j^p) - \frac{\Delta t}{2 \Delta x} (  F_j^p - F_{j-1}^p)
\ee

\end{document}